\documentstyle[12pt]{article}
\newlength{\dinwidth}                       
\newlength{\dinmargin}                      
\def\lsim{\mathrel{\rlap{\lower4pt\hbox{\hskip1pt$\sim$}}
    \raise1pt\hbox{$<$}}}                
\def\gsim{\mathrel{\rlap{\lower4pt\hbox{\hskip1pt$\sim$}}
    \raise1pt\hbox{$>$}}}                
\def\bbar{$B\bar{B}$}
\def\be{\begin{equation}}
\def\ee{\end{equation}}
\def\pts{$p_{sum\; T}^{\;\; 2}$}
\def\chan{$B^-\rightarrow D^{*+}(2010)\pi^-\pi^-\pi^0\rightarrow
(K^-\pi^+\pi^0)\pi^+\pi^-\pi^-\pi^0$}
\begin{document}
\begin{flushright}
BRU/PH/212 \\
hep-ex/9609012 \\
September 1996 \\
\end{flushright}
\vspace*{0.1cm}
\begin{center}  \begin{Large} \begin{bf}
On the Study of \bbar\ Correlations at HERA-B \\  
  \end{bf}  \end{Large}
  \vspace*{5mm}
  \begin{large}
R. Rylko$^a$\\ 
  \end{large}
\end{center}
$^a$ Department of Physics, Brunel, The University of
    West London, Uxbridge UB8 3PH, UK \\
\begin{quotation}
\noindent
{\bf Abstract:}
We analyze the possibility of studying the heavy flavour 
had\-ro\-pro\-duction properties at the HERA-B 
experiment. In addition to the high statistics single inclusive 
$B$\  spectra measurements, the measurement of the  
\bbar\ meson correlations is considered. 
The techniques of momentum estimators, widely used in the charm sector,
are demonstrated to be useful for  the \bbar\ correlation studies
at HERA-B. 
The kinematic limits for the precision of the momentum estimator 
within which the pair spectra can be
measured are determined. The errors are weakly dependent  on the topology
of the multibody $B$\ meson decay.
\end{quotation}
\section{Introduction}
Although the flagship goal of the HERA-B experiment is to study 
the CP violation in the $B$\ system, 
there should be quite an interesting data
sample available for studies of production properties of heavy 
hadrons. 
Here we report on a study of possibilities for measurements of various
heavy hadron production properties with an emphasis on the measurements of 
the distribution of pairs. 
We concentrate on beauty hadrons, as the momentum
estimators were already demonstrated to be a useful tool in the charm sector.
First, we briefly review the existing 
experimental results on the fixed target 
hadroproduction of heavy hadron pairs (mostly for the charm sector) 
and we briefly discuss 
the momentum estimator techniques used. Next, we find the lower boundaries 
for the systematic errors for the distributions of pairs 
when the momentum estimator is used in the kinematic situation of HERA-B.  
This is illustrated for a typical $B$\ decay channel. 
The major systematic effect on the momentum estimator coming from the 
secondary vertex resolution is also evaluated.  
Finally, the errors of the momentum estimator are
studied for a wide range of $B$\ decays classified topologically.

\section{Single inclusive $B$\ production}
With a heavy hadron sample of the order of 10$^5$\ $B$\ mesons per year
\cite{TDR} the HERA-B will be a very competitive 
experiment for the study of
properties of the single inclusive production. 
Although the QCD calculations of the single inclusive heavy hadron production 
have been performed some time ago \cite{LO},
the next-to-leading order (NLO) QCD calculations \cite{NLO} 
must be supplemented
by various soft phenomena \cite{MQCD}
in order to give predictions that agree with
the experimental data. There are two
sources of theoretical uncertainties: the choice of the heavy quark masses, 
the renormalization and factorization scales etc.,
which are input for the NLO QCD calculations,
as well as the 
parameters of the nonperturbative models which supplement them. 
Thus, the resulting total cross-section for the bottom 
production at HERA-B
energies is predicted only within a range of 6-17 nb.  
The single inclusive Feynman-$x$\ and the transverse 
momentum spectra are predicted with similar uncertainties \cite{MQCD}. 

On phenomenological grounds, the high statistics sample of $B$\ mesons may be
used to answer questions concerning the $x_F$\ and $p_T$\ spectra for heavy
hadrons. Namely, several experiments (see e.g.\ \cite{Appel} and references
cited therein) report non-vanishing $x_F^0$, the center of Feynman-$x$\ distribution 
\be
 \frac{d\sigma}{dx_F} \sim (1- |x_F -x_F^0|)^n  \; .
\ee
This offset, although predicted by the theory, 
is experimentally demonstrated only within one standard deviation.
Another problem is the shape (of the tail) of the $p_T$\ spectrum.
While the $e^{-b_2 p_T^2}$\ form is used for most of the data, some
collaborations find the simple exponential form 
$e^{-b_1 p_T}$\  better describes the data, particularly
in the high $p_T$\ region (see e.g.\ \cite{Appel}).  

\section{Current results \bbar\ correlations }

The perturbative and nonperturbative input parameters produce considerable
uncertainties for the predictions of 
the single inclusive heavy hadron production.
The distributions of heavy hadron pairs seems to be
even more sensitive  to the choice of those parameters.
In addition, some nontrivial effects for distributions of pairs,
absent in the leading order QCD (where the heavy hadron pair is 
produced back-to-back), arise entirely as the NLO or
the nonperturbative effects. 
Thus, the distributions of heavy hadron pairs, 
although more difficult
from experimental point of view, 
are an excellent place to study the 
effects of higher order QCD 
and nonperturbative contributions. 

The kinematic variables commonly used to study the hadroproduction of heavy 
hadron pairs may be divided into two groups
\begin{itemize}
\item the angular variables, e.g.\  \\
-  azimuthal angle difference $\Delta \phi=|\phi_1-\phi_2 |$, \\
-  {\it pseudo}rapidity difference $\Delta \eta=|\eta_1 - \eta_2|$;
\item the momentum variables, e.g.\ \\
- the Feynman-$x$\ of a pair $x_F^{pair}$, \\
- the effective mass of the pair $M_{eff}$, \\
- the transverse momentum of the pair \pts$=(\vec{p}_{1T}+\vec{p}_{2T})^2$, \\
- the rapidity difference $y_{diff}=|y_1-y_2|$.
\end{itemize}

A majority of data comes from the fixed target hadroproduction of 
charmed pairs.
The only result for bottom is based on 9 pairs
(not fully reconstructed). 
The results are collected in Table 1.
In all these studies momentum estimators have been used.  

\begin{table}
\begin{center}
\begin{tabular}{c|c|c|c|c|c}
\hline
experiment & No. of & $<p^2_{sum\; T}>$\ & $<M_{eff}>$\ & $<y_{diff}>$\ &
$<\Delta\phi>$ \\ 
& pairs & $[$GeV$^2 ]$ & $[$GeV$]$ & $[1]$ & $[^{o}]$ \\
\hline
E653 p-em. & 35 & -  & $5.56^{+0.21}_{-0.37}$ & 
$1.21^{+0.10}_{-0.13}$ & 
107.1 $\pm$ 9.6 \\
800 GeV \cite{E653} &&&&&\\
WA75 $\pi^- -$em. & 177 & $2.0^{+0.50}_{-0.33}$ & 
$4.59^{+0.14}_{-0.09}$ & 0.80 $\pm$ 0.05
& 109.2 $\pm$ 4.0 \\
350 GeV \cite{WA75} &&&&&\\
WA92 $\pi^- -$Cu & 475  & 1.90 $\pm$ 0.17 & 5.02 $\pm$ 0.16 & - 
& 102.4 $\pm$ 3.6 \\
350 GeV \cite{WA92} &&&&&\\
NA32 $\pi^- -$Cu & 557  & 1.98 $\pm$ 0.11 & 4.45 $\pm$ 0.03 & 0.54 $\pm$ 0.02
& 109.2 $\pm$ 2.4 \\
230 GeV \cite{RR,ACCazi} &&&&&\\
\hline
E653 $\pi^-$-em.  & 9  & 5.0 $\pm$ 2.5 & - & - 
& 147.3 $\pm$ 19 \\
600 GeV \cite{E653bb} & $B\bar{B}$\ &&&&\\
\hline
\end{tabular} 
\caption[junk]{{\it
Experimental results for the heavy hadron momentum correlations. 
The first four are the results for the charm pairs, the last one is for 
the beauty pairs. }}
\end{center}
\end{table}

\section{Momentum estimators}
To study the distributions of angular variables of the heavy hadrons
one needs to know the flight directions of both hadrons. This is 
achieved from the precisely determined  
primary and secondary vertices.
It is enough that one (or even none) of the hadrons is fully 
reconstructed. 

In order to study the momentum correlations, one needs to know
the momenta of both heavy hadrons\footnote{
There is an alternative approach to study the correlations 
of heavy hadron pairs in a 
case of experiment like HERA-B \cite{RYLprog}. In this report
we concentrate on the momentum estimator approach. }.
There are various techniques used for the
estimation of the momentum of decaying heavy hadrons. 
They are based on the precise  
measurement of the position of the primary and  secondary vertex 
and  the measurement of the 
charged decay products' momenta.  
The kinematic techniques described below 
are often supplemented by Monte Carlo information. 
There are two popular approaches:
\begin{itemize}
\item estimator EQ, 
the problem may be solved exactly with three inputs:
the heavy hadron flight vector, the visible decay products'
momenta  and the effective
mass of the invisible decay products.  
\item estimator ET, relies on the assumption that the invisible decay
products' momentum, in the rest frame of the heavy hadron, is perpendicular 
to its flight vector.
Then the corresponding boost from the rest frame may be found to
match the laboratory visible momentum.
\end{itemize} 
The estimator EQ gives the exact answer, which is ambiguous. Then either 
the most probable solution (as suggested by MC or the ET estimator) 
or just the  average of the two solutions may be used. 
For this estimator one needs to
know the effective mass of the invisible decay products. This is simple when 
the invisible decay product is e.g.\ one $\pi^0$, neutrino 
or a narrow state decaying 
into neutrals.  But in a case of e.g.\  nonresonant $\pi^0\pi^0$\ 
in the final state, the necessary estimates of the invisible effective mass 
and the experimental uncertainties of the measured quantities 
needed for the exact solution, limit the applicability of this estimator
to a small class of events. 

The ET estimator is simple to apply and works well for multibody final states
\footnote{{\footnotesize In practical applications,
in order to reduce the error of the estimator, 
the measured distributions are cut 
at the very tail (with a typical loss of a few percent of events).
A strong improvement for the momentum estimator comes from the application
of the cut on the transverse momentum of the visible decay products
with respect to the flight vector. This cut, however, has much stronger
influence on the statistics. For the following 
studies of the errors of the momentum estimator 
we do not optimize the selection cuts, thus keeping more than 
90\%\ of events. } }. 
It has been  used for studies of the single inclusive 
distributions by the E653 collaboration \cite{E653s},
and the charm hadron momentum correlations
in the E653 \cite{E653}, WA75 \cite{WA75}, WA92 \cite{WA92}
and NA32 \cite{RR} data.   

\section{HERA-B and \bbar\ correlations}

According to expectations \cite{TDR},
the HERA-B experiment should collect a sample of about 10$^5$\ 
\bbar\ events per year.
The advantage of the design of the experiment
is that one of the $B$\ hadrons in each event will be fully reconstructed.
The other $B$\ hadron will decay into any decay mode.
The momentum estimator ET has typically a sizable error, however,  
the estimated momentum of the other $B$\ will be added to the well measured,
fully reconstructed $B$. 
Thus, the relative error of the momentum of the \bbar\ pair 
will be roughly halved. 
The results of the simulation \cite{decest},  
with HERA-B parameters and $x_F$\ and $p_T$\
acceptances \cite{TDR}, for the \bbar\ pairs with 
one of the $B$\ mesons fully reconstructed
and the other decaying into \chan\ chain ($D^0$\ decay products are 
in the brackets and both $\pi^0$s are unmeasured) are 
shown in Fig.1. The resolution of the momentum estimator 
for the laboratory momentum of the single $B$\ meson 
is 18 \%, while for
the laboratory momentum of the pair $p^{pair}_{lab}$\ is 9.1\%. 
This results in the errors for the pair 
variables\footnote{The $B$\ mesons in the \bbar\ pair are uncorrelated, thus
giving the averages $< p_{sum\; T}^{\;\; 2} >$=10.3  GeV$^2$\ and
$<M_{eff}>$=11.8 GeV. }  
$\Delta x^{pair}_F$=0.027,
$\Delta p_{sum\; T}^{\;\; 2}$=1.3  GeV$^2$,
$\Delta M_{eff}$= 0.34 GeV and
$\Delta y_{diff}$=0.16. 

\begin{figure}[htb] \label{FIG1}
\vspace{8mm}
\begin{center}
\caption[junk]{{\it
The relative difference of the estimated and the true laboratory 
momentum for the 
single $B$\ meson (left) and for 
the \bbar\ pair (right). One of the $B$\ mesons in the pair
decays into \chan\, with both $\pi^0$s being unmeasured. The
other $B$\ meson is fully reconstructed. }}
\end{center}
\end{figure}

The quoted errors of the pair variables are the lower limits, 
determined by the production and the decay kinematics,  
when the momentum estimator ET is used. 
One of the most important sources of additional systematics is the secondary
vertex resolution\footnote{
The primary vertex resolution at HERA-B is of the
order of 10$\mu$m \cite{TDR} and is neglected here.}.  
To use the momentum estimator, only the flight vector direction, 
and not e.g.\ the decay length, must be known. The longitudinal resolution 
has a little influence
on the flight vector direction. This is further reduced 
if the decay length cut is applied. Indeed,
the simulation with the secondary 
vertex resolution $\sigma_z=500\mu m$ and $\sigma_x=\sigma_y=25\mu m$\ 
\cite{TDR} results in small 
changes of the momentum estimator errors for the pair Feynman-$x$,    
$\Delta x^{pair}_F$=0.028, and for the rapidity difference, 
$\Delta y_{diff}$=0.17, once the cut\footnote{The mean flight path of 
the $B$\ meson at HERA-B energies
is of the order of 9 $mm$.} 
of 3 $mm$\ on the decay length
is applied. The effect is stronger for the transverse momentum  
$\Delta p_{sum\; T}^{\;\; 2}$=1.8  GeV$^2$\
and the effective mass $\Delta M_{eff}$= 0.38 GeV.

\section{Topological approach}

As shown in the previous section, the ET momentum estimator is adequate
within a typical error of $\sim$10\%\ for the 
laboratory momentum of the \bbar\ pair.
The studied decay 
channel has the branching ratio of  (1.5$\pm$0.7)\%\ \cite{RPP}
only. 
On the other hand, the inclusive branching ratios for $B$\ mesons are
$B\rightarrow D^+ X = (26\pm 4)$\%\ and
$B\rightarrow D^0 X = (54\pm 6)$\% \cite{RPP}.
As the $B$\ mesons decay in a large number of channels with small branching
ratios, in any practical approach, one needs to identify the decay vertex of
the other $B$\ hadron and all tracks which belong to this vertex. Exact
knowledge of the decay channel and the decay chain 
is not necessary for the momentum
estimator ET to be applied. 
The dependence on the decay chain for the decays  with the same 
number of tracks emerging
from the secondary vertex is week (see Table 2). 
In addition, the vertex missing 
mass may be used to estimate the number of missing neutrals.
The events may be classified topologically and their systematics
studied for the whole class of events.

\begin{table}
\begin{center}
\begin{tabular}{l|c|c|c|c|c}
\hline
decay & $\Delta p^{pair}_{lab}$ & $\Delta x^{pair}_F$ 
& $\Delta p_{sum\; T}^{\;\; 2}$ & $\Delta M_{eff}$\ & $\Delta y_{diff}$ \\
mode & $[\%]$\ &  & [GeV$^2$] & [GeV] & \\
\hline
$(K\pi)\pi\pi\pi^0\pi^0$ & 8.1 & 0.023 & 1.2 & 0.30 & 0.14 \\
$(K\pi\pi^0)\pi\pi\pi^0$ & 11.1 & 0.034 & 1.4 & 0.43 & 0.20 \\
$(K\pi\pi^0\pi^0)\pi\pi$ & 10.9 & 0.033 & 1.4 & 0.43 & 0.19 \\
\hline
$(K\pi\pi)\pi\pi\pi^0\pi^0$ & 7.5 & 0.021 & 1.2 & 0.28 & 0.13 \\
$(K\pi\pi\pi^0)\pi\pi\pi^0$ & 9.1 & 0.027 & 1.3 & 0.34 & 0.16 \\
$(K\pi\pi\pi^0\pi^0)\pi\pi$ & 9.2 & 0.027 & 1.3 & 0.35 & 0.16 \\
\hline
\end{tabular}
\caption[junk]{{\it
Errors of the momentum estimator ET for various classes of the
$B$\ decaying into the $D^{(*)}+(n\pi)$\ channels.  In brackets are
the decay products of the $D^{(*)}$. 
The $\pi^0$s are not being measured.
The other $B$\ meson
in the event is fully reconstructed. }}
\end{center}
\end{table}

\section{Conclusions }
We show that the HERA-B experiment will provide an excellent opportunity
to study the heavy quark hadroproduction and the perturbative 
and nonperturbative
contributions to this process. It should  resolve long 
standing problems of the single inclusive heavy hadron production 
like the shape of the Feynman-x and $p_T$\ distributions.
The heavy hadron pair distributions are also worth studying
at HERA-B.
Firstly, there should be a large \bbar\ sample for the study of angular 
correlations (comparing with current results). In addition, the
study of the \bbar\ momentum correlations, although difficult, 
is possible with the use of momentum estimators.  
Approximately a few percent of \bbar\ events, with the first $B$\ meson 
fully reconstructed and the second $B$\ meson
partially reconstructed, could be used, 
compared with $\ll 10^{-3}$, when both $B$\ mesons are required 
to be fully reconstructed. The kinematic limits on the 
resolutions of the momentum estimator are of the order of
8-12\%\ for the laboratory momentum of the \bbar\ pair, resulting in
the resolutions 
\begin{itemize}
\item 0.02-0.04 for the $x^{pair}_F$\ distribution, 
\item 1.2-1.4 GeV$^2$\ for the $p_{sum\; T}^{\;\; 2}$\ distribution
(with the average $< p_{sum\; T}^{\;\; 2} >$=10.3  GeV$^2$), 
\item 0.3-0.5 GeV for the $M_{eff}$\ distribution
(with the average $<M_{eff}>$=11.8 GeV),
\item 0.15-0.20 for the $y_{diff}$\ distribution.
\end{itemize}
The source of the main systematic error, namely 
the effects of the secondary vertex resolution, are under control
once the cut on the secondary vertex separation of 
2-3 $mm$\  is applied. 
\vskip1cm
\begin{flushleft}
{\large\bf Acknowledgements}
\vskip0.1cm
\end{flushleft}
I want to thank the workshop organizers for financial support.
My thanks also to Dr.\ R.\ Eichler and Dr.\ A.\ Hasan 
for reading the manuscript and  their helpful remarks.

\end{document}